\documentclass[12pt]{article}  
\usepackage{amsmath}
\usepackage{amssymb} 
\usepackage{euscript}

\usepackage{epsfig}

\usepackage{cite}

\usepackage{alltt}

\usepackage{epic}

\begin{document}   
\allowdisplaybreaks

\begin{titlepage}
\begin{flushright}
{\bf LC-PHSM-2003-068 }\\ {\bf CERN-TH/2003-166} \\{\bf hep-ph/0307331}  \\ {\bf July 2003  }
\end{flushright}
\vspace{1 cm}
\begin{center}{\bf\Large  Probing  the $\cal{CP}$ nature of the Higgs boson 
 }
\end{center}
\begin{center}{\bf\Large  at linear colliders with  $\tau$ spin correlations; }
\end{center}
\begin{center}{\bf\Large the case of mixed scalar--pseudoscalar couplings $^{*}$   }
\end{center}
\vspace{0.8 cm}
\begin{center}
  {\large\bf  K. Desch$^{a}$, A. Imhof $^{b}$, Z. W\c{a}s$^{c,d}$,} 
  ~{\large\bf   M. Worek$^{e}$~  }\\
\vspace{0.2 cm} 
{\em $^a$ Universit\"at Hamburg, Institut f\"ur Experimentalphysik \\
Notkestrasse 85, D-22607 Hamburg, Germany.}\\
{\em $^b$ DESY, Notkestrasse 85, D-22607 Hamburg, Germany.}\\
{\em $^c$ HNINP,
Radzikowskiego 152, 31-342 Cracow, Poland.}\\   {\em $^d$ CERN, Theory
Division, CH-1211 Geneva 23, Switzerland.}\\   {\em $^e$ Institute
of Physics, University  of Silesia\\ Uniwersytecka 4,    40-007
Katowice, Poland.}
\end{center}
\vspace{4mm}
\begin{abstract}

The prospects for the measurement of the pseudoscalar admixture in the
$h\tau\tau$ coupling to a Standard Model  Higgs boson  of 120 GeV mass
are discussed in a quantitative  manner for $e^+e^-$ collisions of 350
GeV center-of-mass energy.  Specific angular distributions in the
$h\to \tau^+\tau^-,\; \tau^\pm\to \rho^\pm
\bar{\nu}_{\tau}(\nu_{\tau})$  decay chain  can be used to probe
mixing angles of  scalar--pseudoscalar $h\tau\tau$ couplings.  In the
discussion of the feasibility of the method, assumptions on the
properties of a future detector for an $e^+e^-$ linear collider such
as TESLA are used. The Standard Model Higgsstrahlung production
process is taken as an example.   For the expected performance of a
typical Linear Collider set-up, the sensitivity of a measurement of
the scalar--pseudoscalar mixing angle turned out  to be 6$^{\circ}$.
It will be straightforward  to apply our results to estimate the
sensitivity  of a measurement, in cases  another scenario of the
Higgs boson sector (Standard Model  or not) is chosen by nature. The
experimental error of the method is expected to be  limited by the
statistics.

\end{abstract}
\begin{center}
{\it Published in ~Phys. Lett. B579 (2004) 157-164.  }
\end{center}

\vspace*{1mm}
\bigskip
\footnoterule

\noindent
{\footnotesize
\begin{itemize}
\item[${*}$]
This work is partly supported by
the Polish State Committee for Scientific Research
(KBN) grants Nos 2P03B00122, 5P03B09320
and the European Community's Human Potential
Programme under contract HPRN-CT-2000-00149 Physics at Colliders.
\end{itemize}
}

\end{titlepage}

\section{Introduction}

The transverse spin effects in $\tau$ pair production can be helpful
to distinguish between the scalar $\mathcal{J^{PC}}=0^{++}$ and
pseudoscalar $\mathcal{J^{PC}}=0^{-+}$ natures of the spin zero
(Higgs) particle, once it is discovered in future accelerator
experiments.  To address resolution issues, it is necessary  to
perform Monte Carlo studies,   where the significant details of
theoretical effects and detector conditions can   be included.   To
enable such studies  we have extended the algorithm    of
Refs. \cite{Pierzchala:2001gc,Worek:2001hn} of the {\tt TAUOLA}
$\tau$-lepton decay library
\cite{Jadach:1990mz,Jezabek:1991qp,Jadach:1993hs}  to include the
complete spin effects of $\tau$ leptons originating  from the spin
zero  particle.

In Refs. \cite{Was:2002gv,Worek:2003zp} the reaction chain   $e^+e^-
\to Z (H/A^{0})$,     $H/A^{0} \to \tau^{+}\tau^{-}$,   $\tau^{\pm}\to
\pi^{\pm}\bar{\nu}_{\tau}({\nu}_{\tau})$ was    studied. It was found
that even small effects of smearing   seriously deteriorate the
measurement resolution.  However, using the {\tt TAUOLA} spin
interface, we devised a very promising method for the measurement of
the Higgs boson parity, see  Ref.~\cite{Bower:2002zx}.  It turns out
that the  spin effects of the decay chain $H/A^{0} \to
\tau^{+}\tau^{-} \to \rho^{+}\bar{\nu}_{\tau}\rho^{-}\nu_{\tau} \to
\pi^{+}\pi^0\bar{\nu}_{\tau}\pi^{-}\pi^0\nu_{\tau}$  give a parity test
independent of both model (e.g. SM, MSSM) and Higgs boson
production  mechanism (e.g. Higgsstrahlung, WW fusion).
In the rest frame of the $\rho^{+}\rho^{-}$ system we defined the
acoplanarity angle $\varphi^{*}$ as the one between the two planes
spanned by    the immediate decay products (the $\pi^\pm$ and $\pi^0$)
of the two $\rho$'s.  This angular distribution of the $\tau$ decay
products, which is sensitive to the Higgs boson parity,  once
additional selection cuts are applied, is measurable using typical
properties of a future detector at an $e^+e^-$ linear collider.  Using
reasonable assumptions about the SM production cross section and about
the measurement resolutions we have found that,   with 500 fb$^{-1}$
of luminosity at a  500 GeV $e^+e^-$ linear collider, the $\mathcal
{CP}$ of a 120  GeV Higgs boson can be measured to a confidence level
greater     than 95$\%$.

In Ref.~\cite{Desch:2003mw} we  demonstrated that a measurement of the
$\tau$ impact parameter  in one-prong $\tau$ decay is useful for the
determination of the  Higgs boson parity in the
$H/A^{0}\to\tau^{+}\tau^{-}$;
$\tau^{\pm}\to\rho^{\pm}\bar{\nu}_{\tau}(\nu_{\tau})$ decay chain. We
estimated that, for a detection set-up such as  TESLA, use of the
information from the $\tau$ impact parameter can improve the
significance of the measurement of the parity of a Standard Model
$120$ GeV Higgs boson to $\sim$ 4.5$\sigma$ and in general  by a
factor of about 1.5 with respect to the method where this information
is not used.  So far we have not exploited the possibility of using
decay modes other than $\tau^{\pm} \to \rho^{\pm}
\bar{\nu}_{\tau}(\nu_{\tau})$.  Additional  modes are expected to
further increase the separation power.

In this paper we study the more general case where mixed scalar and
pseudoscalar couplings of the  Higgs boson to $\tau$ leptons are
simultaneously allowed, see e.g. Ref.~\cite{Abe:2001np}.

Our paper is organized as follows:  In Section 2 we present basic
properties of the density matrix for the pair  of $\tau$ leptons
produced in Higgs boson decay. In Section 3 we define  our observable
and in Section 4 our Monte Carlo set-up. Our results are presented in
Sections 5 and 6, first with an idealized detector set-up and then
with more realistic assumptions on the detector and integrated
luminosity. A summary, Section 7, closes the paper.


\section{Spin weight for the mixed scalar--pseudoscalar case}

Let us here,  only very briefly describe the basic properties of the
spin correlations and their implementation in our Monte Carlo
algorithm.   We will not repeat the detailed description of the method
(which can be found in Ref.~\cite{Jadach:1990mz})  or the algorithm
(which is given in Ref.~\cite{Was:2002gv}). We will discuss the points
necessary to understand  the case  of mixed scalar--pseudoscalar
coupling of $h\tau \tau$.

The main  spin weight of our algorithm for generating the physical
process of $\tau$ lepton  pair production in Higgs boson decay, with
subsequent decay of $\tau$ leptons as well,  is given by
\begin{equation}
wt=\frac{1}{4}\left(1+\sum^{3}_{i=1}\sum^{3}_{j=1} R_{ij}h_{1}^{i}
h_{2}^{j}\right),
\end{equation}
where $h_{1}$ and $h_{2}$  are the polarimeter vectors  that depend
respectively on $\tau^{\pm}$ decay products momenta; $R_{ij}$ is the
spin density  matrix. For the mixed scalar--pseudoscalar case, when the
general Higgs  boson Yukawa coupling to the $\tau$ lepton
\begin{equation}
\label{coupl}
\bar{\tau}(a+ib\gamma_{5})\tau
\end{equation}
is assumed, we get the following non-zero components of $R_{ij}$:
\begin{equation}
R_{33}=-1,~~~~~
R_{11}=R_{22}=\frac{a^{2}\beta^{2}-b^{2}}{a^{2}\beta^{2}+b^{2}},~~~~~~~
R_{12}=-R_{21}=\frac{2ab\beta}{a^{2}\beta^{2}+b^{2}},
\end{equation}
where $\beta=\sqrt{1-\frac{4m^{2}_{\tau}}{m^{2}_{H}}}$.  If we express
Eq.~(\ref{coupl}) with the help of the scalar--pseudoscalar mixing
angle $\phi$:
\begin{equation}
\label{coupla}
\bar{\tau}N(\cos\phi+i\sin\phi\gamma_{5})\tau,
\end{equation}
the  components of the spin density matrix can be expressed in the
following way:
\begin{equation}
R_{11}=R_{22}=\frac{\cos\phi^{2}~\beta^{2}-
\sin\phi^{2}}{\cos\phi^{2}~\beta^{2}+\sin\phi^{2}}, ~~~~~~~
R_{12}=-R_{21}=\frac{2\cos\phi \sin\phi~\beta}
{\cos\phi^{2}~\beta^{2}+\sin\phi^{2}}.
\end{equation}
In  the limit $\beta \to 1$ these expressions reduce to the components
of the rotation  matrix for the rotation around the $z$ axis by an
angle $-2\phi$:
\begin{equation}
R_{11}=R_{22}=\cos2\phi, ~~~~~~~ R_{12}=-R_{21}=\sin2\phi.
\end{equation}

The Higgs boson  parity information must be extracted from   the
correlations between $\tau^{+}$ and $\tau^{-}$ spin components, which
are   further reflected in correlations between     the $\tau$ decay
products in the plane transverse to the    $\tau^{+}\tau^{-}$
axes. The same will now apply  to  the mixing scalar--pseudoscalar
case. To better visualize the effect to be measured, let us write  the
decay probability for the mixed scalar--pseudoscalar  case, using the
conventions of Ref.~\cite{Kramer:1994jn}:
\begin{equation}  
\Gamma(h_{mix}\to \tau^{+}\tau^{-}) \sim 1-s^{\tau^{+}}_{\parallel}
s^{\tau^{-}}_{\parallel}+ s^{\tau^{+}}_{\perp}
R(2\phi)~s^{\tau^{-}}_{\perp},
\label{densi}  
\end{equation}   
where $R(2\phi)$ can be understood as  an operator for the rotation by
an angle $2\phi$  around the ${\parallel}$ direction.  The
$s^{\tau^{-}}$ and $s^{\tau^{+}}$ are  the $\tau^\pm$ polarization
vectors, which are defined    in their respective rest frames.  The
spin quantization axes are oriented in the $\tau^{-}$ flight
direction.     The symbols ${\parallel}$/${\perp}$ denote components
parallel/transverse    to the Higgs boson momentum as seen from the
respective $\tau^\pm$  rest frames.

It is straightforward to see that the pure scalar case  is reproduced
for $\phi=0$. Then $R_{11}=+1$, $R_{22}=+1$ and $R_{33}=-1$ are
obtained, and  the limit $\beta \to 1$ does not need to be taken. For
$\phi=\pi/2$ we reproduce  the pure pseudoscalar case. We get
$R_{11}=-1$, $R_{22}=-1$ and $R_{33}=-1$. Also in this case, the
$\beta \to 1$ limit was not needed.


\section{The acoplanarity of the $\rho^+$ and $\rho^-$ decay products}
 
To  facilitate reading,  let us recall here some elements of the
observables that were presented  in
Refs.~\cite{Bower:2002zx,Desch:2003mw} and can be used to measure the
Higgs boson parity. We will stress only those points that required
modification.  The method relies on measuring the acoplanarity angle
of the two planes, spanned on  $\rho^{\pm}$ decay products and defined
in the $\rho^{+}\rho^{-}$ pair rest frame.  For that purpose the
four-momenta of $\pi^{\pm}$ and $\pi^{0}$ need to be reconstructed
and, combined, they will  yield the $\rho^{\pm}$ four-momenta.  All
reconstructed four-momenta are then boosted into the
$\rho^{+}\rho^{-}$ pair rest frame. The acoplanarity angle
$\varphi^{*}$, between the planes of the $\rho^{+}$ and $\rho^{-}$
decay products is defined in this frame.  In the previous papers only
the range $0 <\varphi^{*} < \pi$ was interesting  and thus
reconstructed, as this was sufficient to distinguish between two
possibilities: scalar or pseudoscalar Higgs boson, differing by the
sign of the transverse spin correlation.   The angle was defined with
the help of its cosine and with the help of the two  vectors ${\bf
n}_\pm$ normal to the  planes namely ${\bf n}_\pm= {\bf p}_{\pi^\pm}
\times  {\bf p}_{\pi^0}$, $\cos \varphi^{*} =\frac{ {\bf n}_+ \cdot
{\bf n}_-}{|{\bf n}_+| |{\bf n}_-|} $.

\begin{figure}[!ht]
\begin{center}  
\epsfig{file=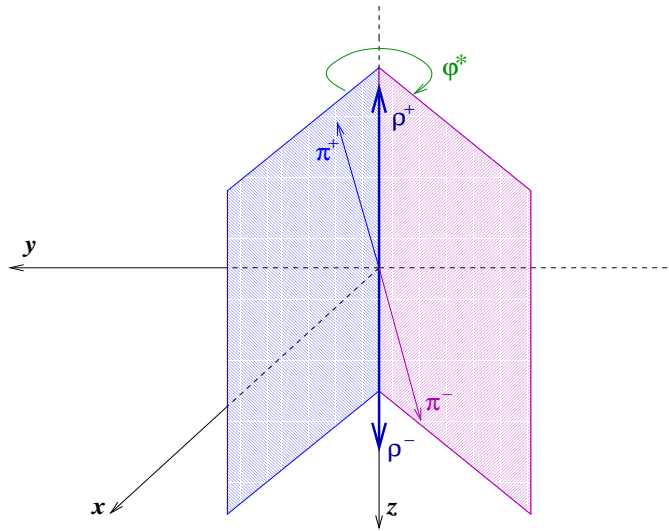,width=90mm,height=70mm}
\end{center} 
\caption  
{\it Definition of the  $\rho^{+}\rho^{-}$ decay products'
acoplanarity distribution angle $\varphi^{*}$, in the rest frame  of
the $\rho^{+}\rho^{-}$ pair. The  range for   $\varphi^*$ is  $0 \le
\varphi^* \le 2\pi$. Note that, for better visualization, we use in
this figure the momenta of $\pi^\pm$ and $\rho^\pm$ (rather than $\pi^0$'s
from $\rho^\pm$ decays) to define the planes. The two ways of
defining the planes are equivalent if no  reconstruction errors are
taken.}
\label{observabla}  
\end{figure}

For the present use, such a definition is insufficient. As can be seen
from Eq.~(\ref{densi}) the correlation, in the case of the Higgs boson
of combined scalar and pseudoscalar couplings of Eq.~(\ref{coupla})
and the mixing angle $\phi$, is between transverse components of
$\tau^+$ spin polarization vector and transverse components of
$\tau^-$ polarization vector  {\it rotated by an angle
$2\phi$}. Therefore  the full range of the variable $0 <\varphi^{*} <
2\pi$ is of  physical  interest. To distinguish between the two cases
$\varphi^{*}$ and  $2\pi-\varphi^{*}$ it is sufficient, for example,
to find the sign of    
$p_{\pi^-} \cdot {\bf n}_+$. When it is negative, the angle
$\varphi^{*}$ as defined above (and  in the range $0 <\varphi^{*} <
\pi$) is used.  Otherwise it is replaced by $ 2\pi - \varphi^{*}$.  If
no separation was made, the parity effect, in case of mixed
$h\tau\tau$ coupling, would wash itself out (see Fig.~\ref{rys1},
later in the text).  For the graphical representation of the
definition of the angle $\varphi^{*}$, see Fig.~\ref{observabla}. The
figure visualizes  the relation between the observable and
Eq.~(\ref{densi}) as well.

Additional selection cuts  need to be applied. Otherwise the acoplanarity
distribution is not sensitive to transverse spin effects (and thus to
Higgs boson parity) at all.  The events need to be  divided into two
classes, depending on the sign of  $y_{1}y_{2}$, where
\begin{equation}
y_1={E_{\pi^{+}}-E_{\pi^{0}}\over E_{\pi^{+}}+E_{\pi^{0}}}~;~~~~~
y_2={E_{\pi^{-}}-E_{\pi^{0}}\over E_{\pi^{-}}+E_{\pi^{0}}}.
\label{y1y2}
\end{equation}
The energies of $\pi^\pm,\pi^0$ are to be taken in the  respective
$\tau^\pm$ rest frames. In Refs.~\cite{Bower:2002zx,Desch:2003mw} the
methods of reconstruction of the replacement $\tau^\pm$ rest frames were
proposed with and without the help of the $\tau$ impact parameter.  We
will use these methods here as well, without any modification.


\section{The Monte Carlo}

If any non-zero CP-odd admixture to the Higgs is present, not only is the
distribution of the Higgs decay products modified, but also the 
distribution of its production 
angle~\cite{Abe:2001np,Kramer:1994jn,Grzadkowski:1995rx}.
In this paper, we simulate production angular distributions as in the
SM, but this assumption has no influence on the validity of the analysis.
In particular, the detection efficiencies for pure CP-even and pure CP-odd
Higgs bosons do not differ significantly. In order to study the
sensitivity of $h \to\tau^+\tau^-$ observables, we assume a production
rate independent of the size of the CP-odd admixture, i.e.~the SM
production rate of a CP-even Higgs. 

  The production process $e^+e^- \to Z h \to \mu^+ \mu^- (q\bar q) \tau^+ \tau^- $ 
has been chosen, as an representative example, 
and simulated with the Monte Carlo program
{\tt PYTHIA 6.1} \cite{Sjostrand:2000wi}. The Higgs boson mass of 120 GeV and 
a    center-of-mass energy of 350 GeV was chosen. 
The effects of initial state    bremsstrahlung were
included.
 For the sake of our
discussion and in all of our  samples the $\tau$ decays  have been
generated with the {\tt TAUOLA}   Monte Carlo library
\cite{Jadach:1990mz,Jezabek:1991qp,Jadach:1993hs}.  As usual, to
facilitate the interpretation of the results, bremsstrahlung effects
in decays were not taken into account. Anyway, with the help of
additional simulation, we have found this effect to be rather small.
To include the full spin effects in the    $h \to \tau^+\tau^-$,
$\tau^{\pm} \to \rho^{\pm}\bar{\nu}_{\tau}(\nu_{\tau})$,
$\rho^{\pm}\to\pi^{\pm}\pi^{0}$  decay  chain, the   interface
explained in Ref.~\cite{Was:2002gv} was used, with the extensions
discussed in Section 2.


\section{Idealized results} 


\subsection{Resolution parameters}  

To test the feasibility of the measurement, some assumptions  about
the detector  effects had to be made. We include, as the most critical
for our discussion, effects due to    inaccuracies in the measurement
of the   $\pi^\pm,\pi^0$ momenta and of the $\tau^\pm$ leptons impact
parameters.    We assumed Gaussian spreads of the measured quantities
with    respect to the generated ones and we used the following
algorithm to reconstruct the energies of $\pi$'s in their respective
$\tau^\pm$ rest frames, exactly as in the case of the studies
presented in Refs.~\cite{Bower:2002zx,Desch:2003mw}.
  
\begin{enumerate}  
\item    {\bf Charged-pion momentum:}   We assume a 0.1\% spread on
its energy and direction.
\item    {\bf Neutral-pion momentum:}   We assume an energy spread of
$ 5 \% \over \sqrt{E [{\rm GeV}]}$. For the $\theta$ and $\phi$
angular spread we assume  $ {1 \over 3}  {2 \pi \over 1800}$.  These
$\pi^{0}$ resolutions can be    achieved with a 15\% energy error and
a 2$\pi$/1800 direction error in the gammas   resulting from the
$\pi^0$ decays.  These resolutions have been approximately verified
with {\tt SIMDET}~\cite{Pohl:2002vk}, a parametric Monte Carlo program
for  TESLA detector~\cite{Behnke:2001qq}, as well as with other
studies, see e.g. Refs.~\cite{evtsel1,evtsel2}.
\item {\bf The reconstructed Higgs boson rest frame:} We assume a
spread of $2$ GeV with respect to the transverse momentum of the
reconstructed Higgs boson momentum,  and  $5$ GeV for the longitudinal
component, to mimic the  beamstrahlung effect.
\item {\bf The impact parameter: } The angular resolution of the
$\tau$ impact parameter  has been simulated for a TESLA-like
detector. The simulation is based on the anticipated performance of a
5-layer CCD vertex detector, as described in
Ref.~\cite{Behnke:2001qq}. For Higgsstrahlung events with
$h_{SM}\to\tau^+\tau^-$ and $\tau^\pm \to
\rho^\pm\bar{\nu}_\tau(\nu_{\tau})$ at $m_{h_{SM}} = $ $120$  GeV and
$\sqrt{s} = $ $350$ GeV, the angular resolution has been found%
~\cite{Desch:2003mw} to be approximately 25$^\circ$.
\end{enumerate}  


\subsection{Numerical results} 

We have used the scalar--pseudoscalar mixing angle
$\phi=\frac{\pi}{4}$ and,  as the reference, we have used the  pure
scalar case $\phi=0$.

\begin{figure}[!ht]
\begin{center} 
\epsfig{file=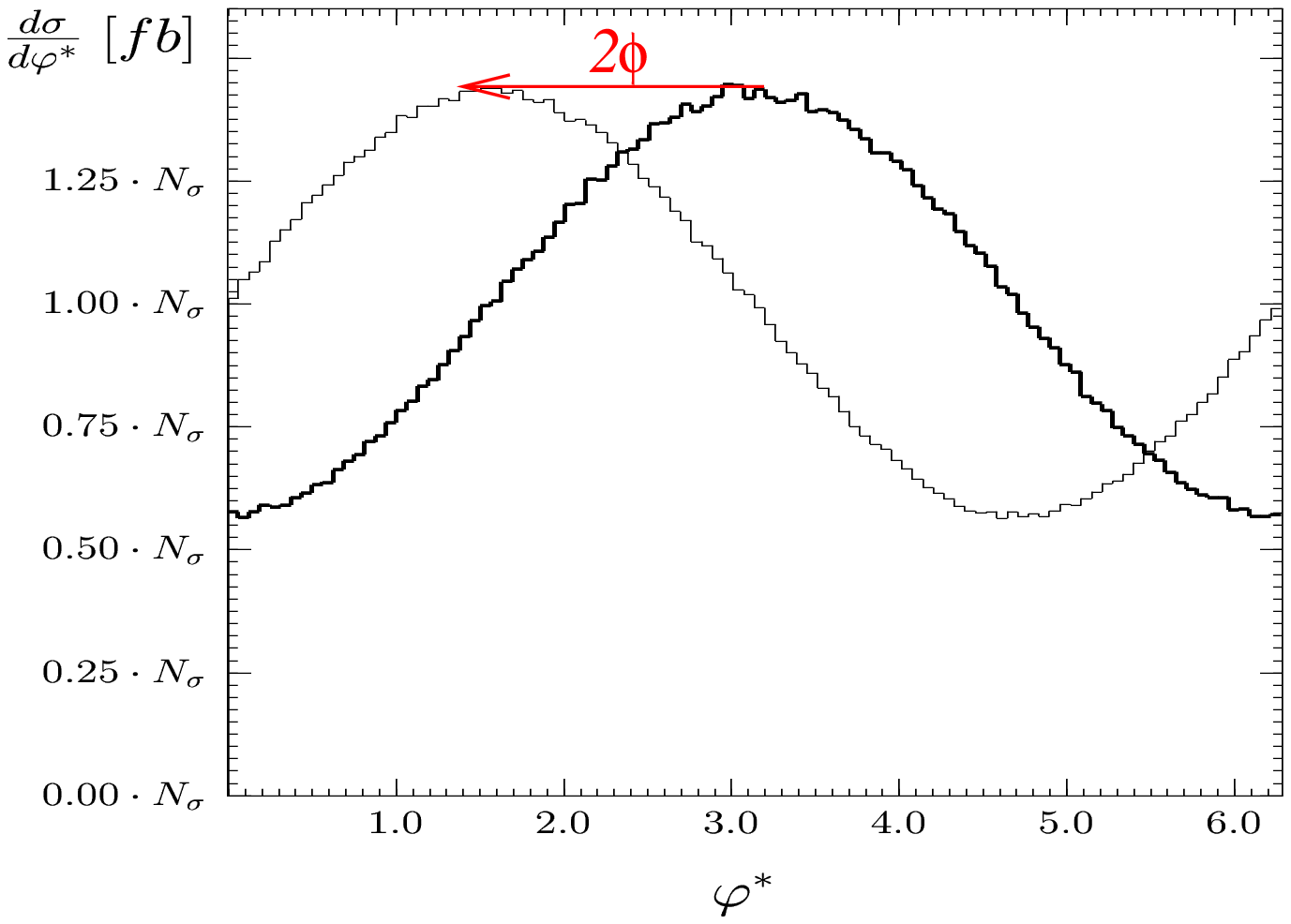,width=75mm,height=70mm}
\epsfig{file=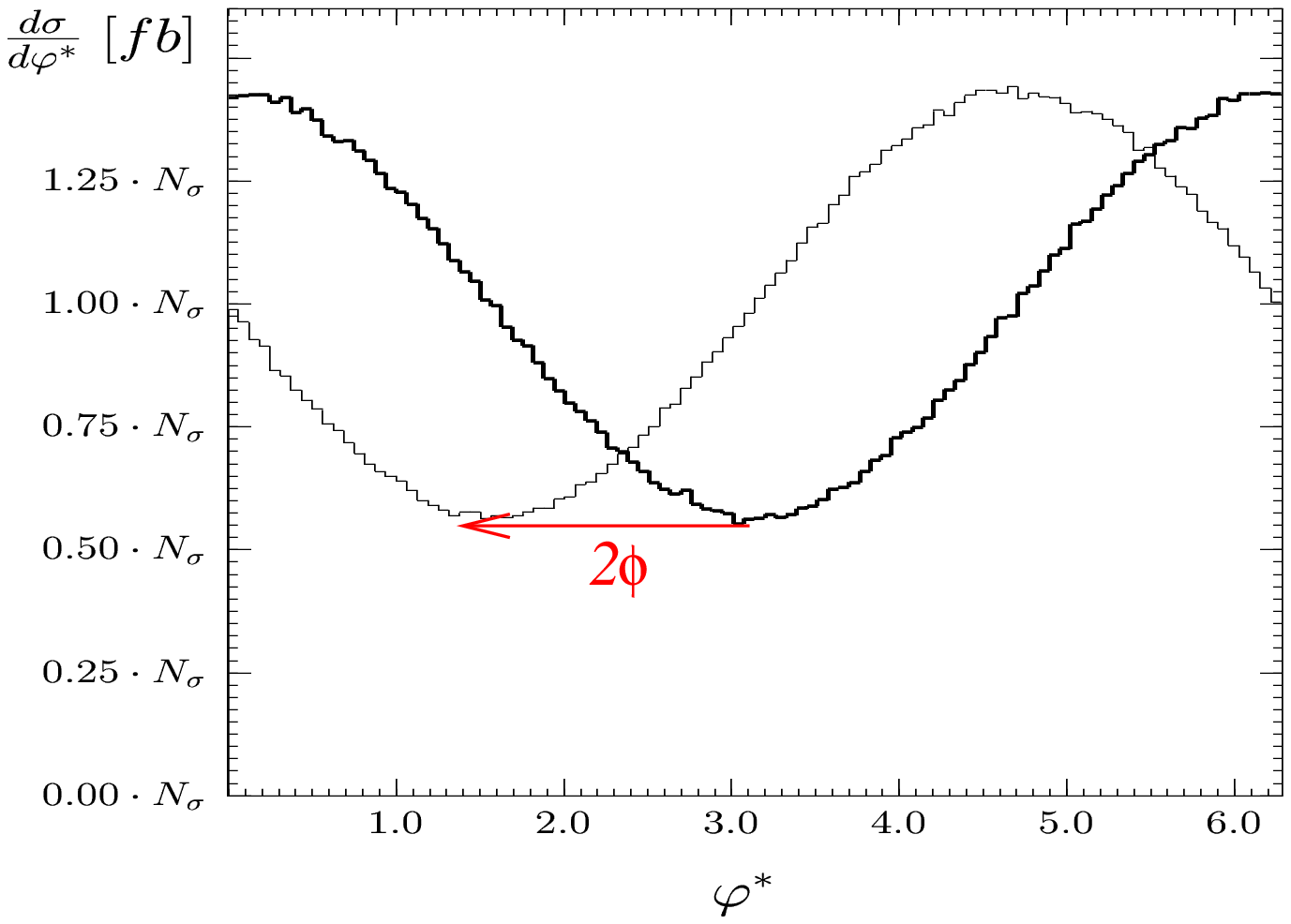,width=75mm,height=70mm}
\end{center} 
\caption  
{\it  The acoplanarity distribution  (angle $\varphi^{*}$) of the
$\rho^+ \rho^-$ decay products     in the rest frame of the $\rho^+
\rho^-$ pair.  Gaussian smearing of $\pi$'s momenta as described
in Section 5.1 is included. However, generator level
$\tau^\pm$ rest frames are used.  The thick line corresponds to a
scalar Higgs boson,  the thin line to a mixed one.   The left figure
contains events with $y_1 y_2 > 0$, the right one is for $y_1 y_2 < 0$.
In our paper, that is for the  350 GeV $e^+e^-$ CMS  
(scalar 120 GeV mass)
 Higgsstrahlung production we took
$N_\sigma=62.7 \cdot 10^{-3}\; [fb]$ for the scale of the plot. 
In general case 
$N_\sigma=\frac{1}{4\pi}\sigma_{total}(e^+e^- \to XH) {\cal BR}(H \to
\tau^+\tau^-) \bigl( {\cal BR}(\tau \to \rho \nu_\tau) \bigr)^2 $
is a suitable choice.
}
\label{rys1}  
\end{figure}

\begin{figure}[!ht]
\begin{center} 
\epsfig{file=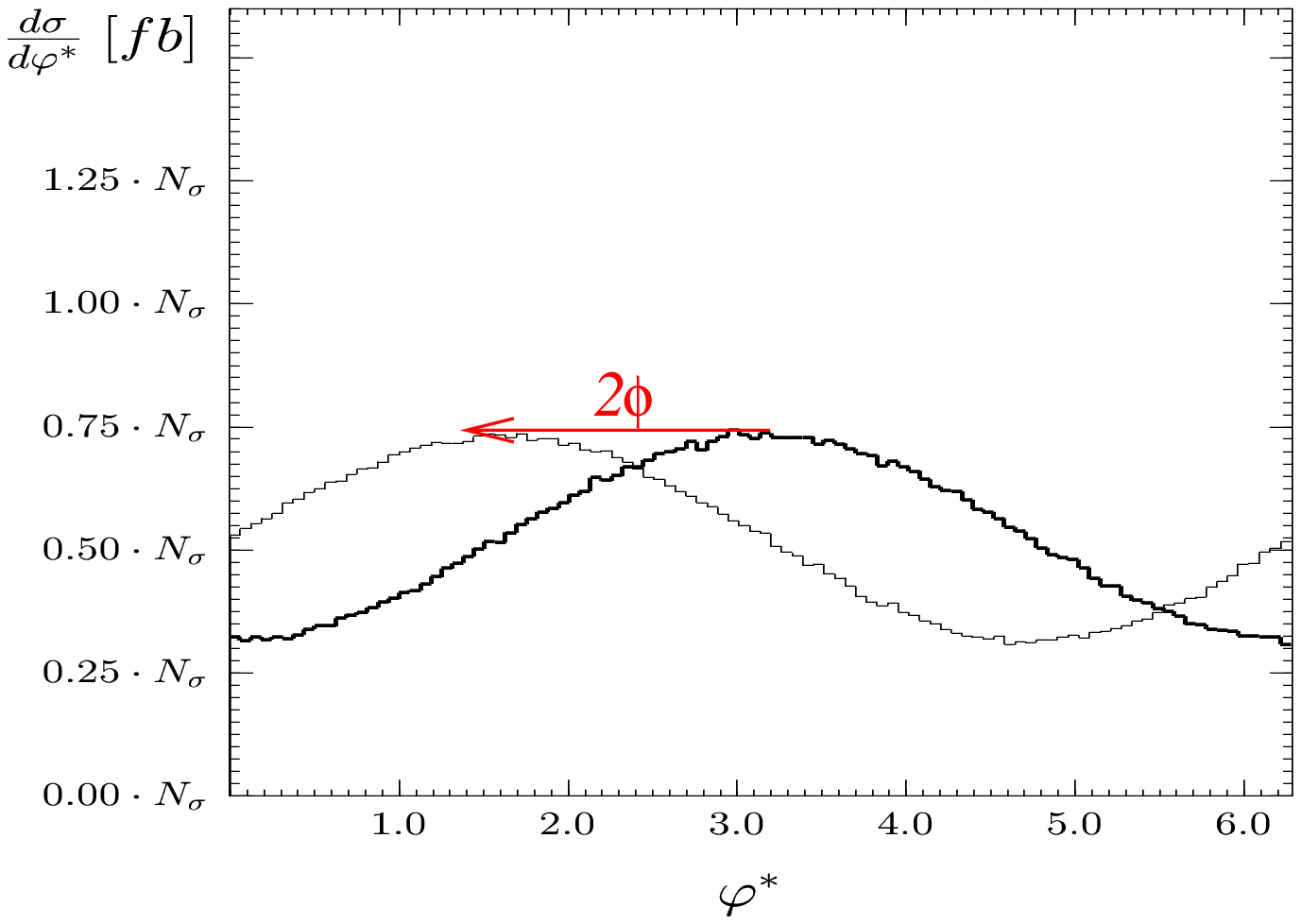,width=75mm,height=70mm}
\epsfig{file=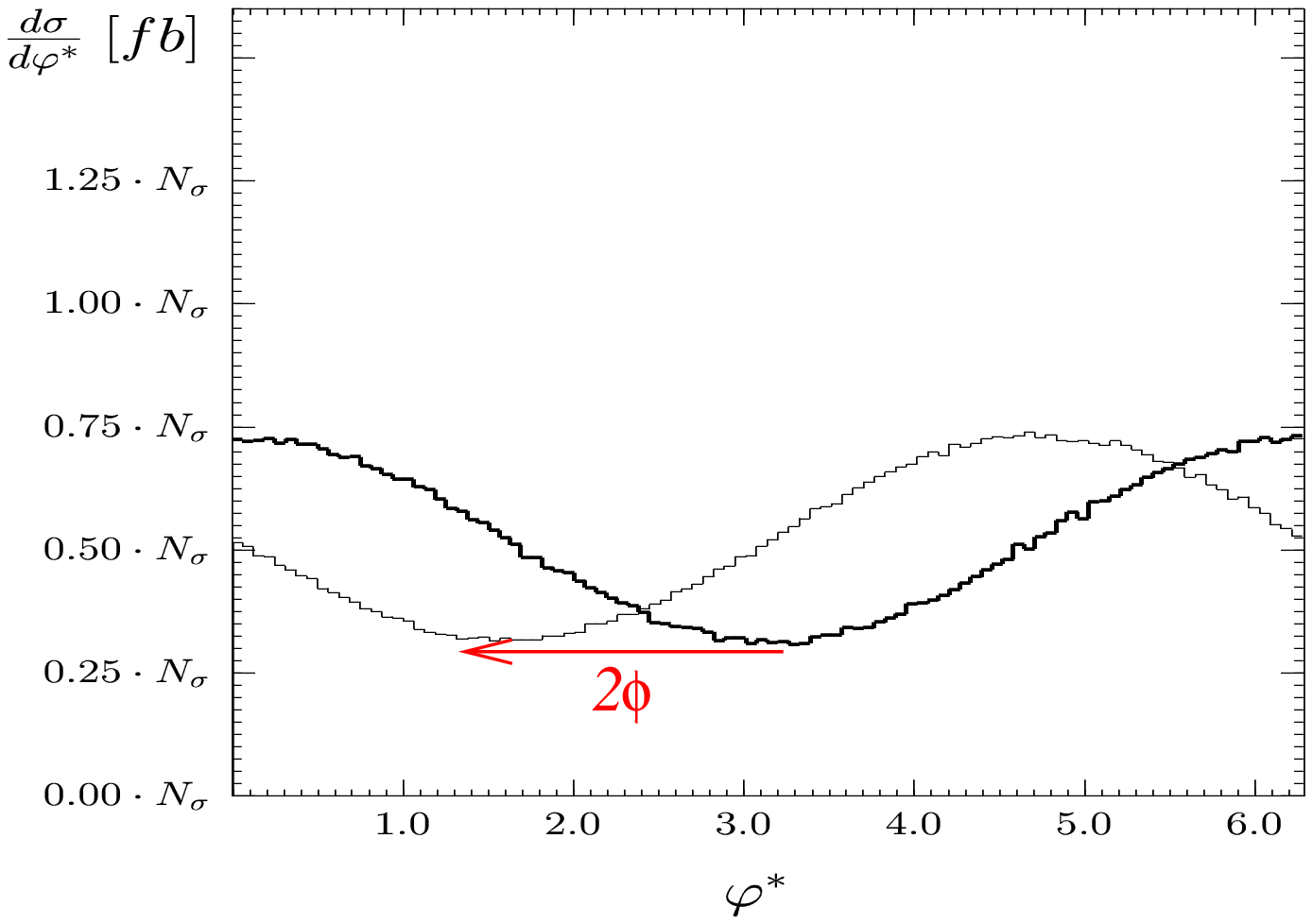,width=75mm,height=70mm}
\end{center} 
\caption  
{\it  The acoplanarity distribution  (angle $\varphi^{*}$) of the
$\rho^+ \rho^-$ decay products     in the rest frame of the $\rho^+
\rho^-$ pair. Gaussian smearing of $\pi$'s  and Higgs boson momenta,
 as described
in Section 5.1 is included. Only events where the signs
of the  energy differences $y_1$ and $y_2$ are the same, if calculated
using the method described in Ref.~\cite{Bower:2002zx} and if
calculated  with the help of the $\tau$ impact parameter
Ref.~\cite{Desch:2003mw}, are taken.  The thick line corresponds to a
scalar Higgs boson, the thin line to a mixed one.  The left figure
contains events with $y_1 y_2 > 0$, the right one is for $y_1 y_2 <
0$. 
In our paper, that is for the  350 GeV $e^+e^-$ CMS  
(scalar 120 GeV mass)
 Higgsstrahlung production we took
$N_\sigma=62.7 \cdot 10^{-3}\; [fb]$ for the scale of the plot. 
In general case 
$N_\sigma=\frac{1}{4\pi}\sigma_{total}(e^+e^- \to XH) {\cal BR}(H \to
\tau^+\tau^-) \bigl( {\cal BR}(\tau \to  \rho \nu_\tau) \bigr)^2 $
is a suitable choice.
}
\label{rys4}  
\end{figure}

In Fig.~\ref{rys1} the acoplanarity distribution  angle $\varphi^{*}$
of the $\rho^+ \rho^-$ decay products which was defined  in the rest
frame of the reconstructed $\rho^+ \rho^-$ pair, is shown.
Unobservable generator-level   $\tau^{\pm}$ rest frames are used for
the calculation of selection cuts.  The two plots represent events
selected by the differences of $\pi^\pm\pi^0$ energies, defined in
their respective $\tau^\pm$ rest frames. In the left plot, it is
required that $y_1 y_2 > 0$, whereas in the right one, events with
$y_1 y_2 < 0$ are taken. This figure quantifies the size of the parity
effect in an idealized condition, which we will attempt to approach
with realistic ones.  The size of the effect was substantially
diminished when a detector-like set-up was included for $\tau^\pm$
rest frames reconstruction as well,  see Fig.~\ref{rys4}, in exactly
the same proportion as in Ref.~\cite{Bower:2002zx}. The  general shape
of the distributions remained.

At the cost of introducing cuts, and thus reducing the number of
accepted events, we could achieve some improvement of the method, as
in Ref.~\cite{Desch:2003mw}.  If we require the signs of the
reconstructed energy differences $y_1$ and $y_2$ (Eq.~(\ref{y1y2})) to be
the same whether the method is used with or without the help of the
$\tau$  lepton impact parameter,  only  $\sim 52\%$ of events are
accepted.  The relative size of the parity effect increases. Results
are presented in Fig.~\ref{rys4}.

\section{Simulation with detector effects}

 In order to assess the possibilities for a measurement of the
acoplanarity distribution described in Section~2, we perform a
detailed simulation of Higgs bosons produced in the Higgsstrahlung
process using {\tt PYTHIA~6.2}~\cite{Sjostrand:2000wi} for the
production process and the modified version of {\tt TAUOLA} described
above to  generate samples of signal events. These events are then
passed  through a simulation of the TESLA detector ({\tt
SIMDET}~\cite{Pohl:2002vk}) accounting for the acceptance and
anticipated resolution of the tracking devices and calorimeters
corresponding to the detector proposed in the   TESLA
TDR~\cite{Behnke:2001qq}.

\begin{figure}[!ht]
\begin{center} 
{\epsfig{file=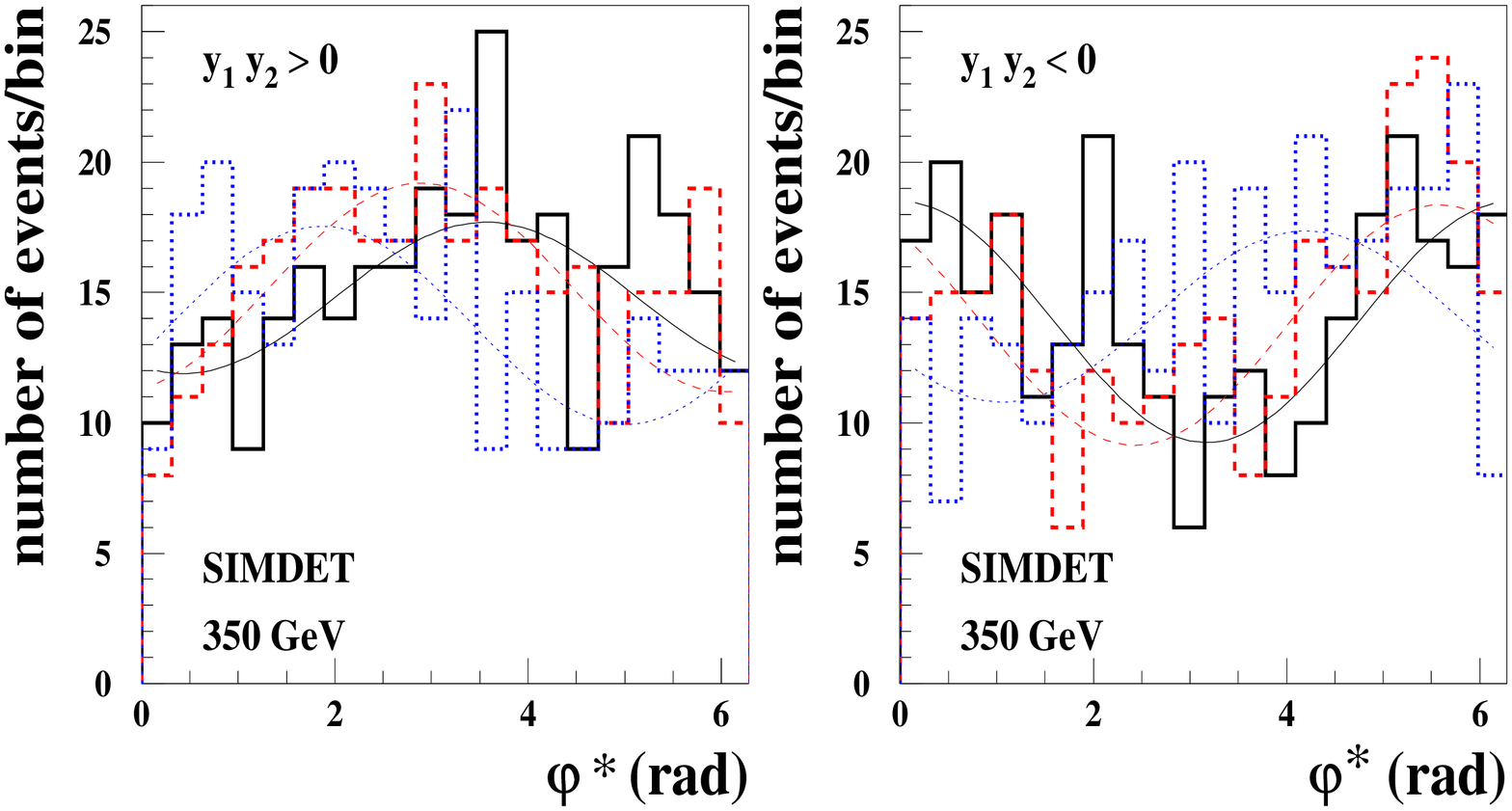,width=160mm,height=70mm}}
\end{center} 
\caption  
{\it Distribution of the reconstructed acoplanarity angle $\varphi^*$
for $\phi = 0$ (full histogram), $\phi = \pi/8$ (dashed histogram) and
$\phi = \pi/4$ (dotted histogram) for $y_1 y_2 > 0$ (left) and $y_1
y_2 < 0$ (right). The lines indicate the results of the corresponding
fits (see text).}
\label{aco-shape-2ab-rec}
\end{figure}

Signal samples \footnote{ Note that this integrated luminosity is
larger by a factor of 2 than the one used in
Refs.~\cite{Bower:2002zx,Desch:2003mw} to estimate the sensitivity of
our Higgs boson parity observable. Also, the Higgstrahlung production
cross section (see  e.g. \cite{Aguilar-Saavedra:2001rg})  is more than
2 larger at 350 GeV than at 500 GeV center-of-mass energy. On the
other hand,  here we do not use   the information from the $\tau$
impact parameter, which can be  useful to improve the sensitivity of a
measurement of the mixing angle $\phi$.}   of 1~ab$^{-1}$ at 350~GeV
center-of-mass energy were generated for scalar--pseudoscalar mixing
angles $\phi = 0$, $\pi/8 $ and $\pi/4 $.  With detector simulation
the $\tau$ leptons decaying to $\pi^\pm \pi^0$ from Higgs decays were
reconstructed as isolated  jets with only one charged track (the
reconstructed $\pi^\pm$)  and additional neutral clusters (the
reconstructed $\pi^0$). The  $\pi^\pm$ and $\pi^0$ momenta were
combined to form a reconstructed $\rho^\pm$.  The acoplanarity angle
$\varphi^{*}$ was calculated in the reconstructed $\rho^+\rho^-$ rest
frame. Two event classes are formed according to the sign of $y_1
y_2$, where $y_1$ and $y_2$ are calculated in the laboratory
frame. The resulting $\varphi^{*}$ distributions for the three  $\phi$
cases are  shown in Fig.~\ref{aco-shape-2ab-rec} as histograms, each
containing  about 0.5~ab$^{-1}$ statistics.

To extract the scalar--pseudoscalar mixing angles $\phi$ the functions
$a*\cos(\varphi^{*} -2\phi) + b$ (for $y_1 y_2 > 0$) and
$a*\cos(\varphi^{*}- 2\phi + \pi) + b$ (for $y_1 y_2 < 0$) were used
to fit 2$\phi$ to the reconstructed  acoplanarities $\varphi^{*}$,
gained from simulated detector signals. The constants $a$ and $b$ were
additional free variables of the 3-parameter fit.  The resulting
functions are also shown as lines  in Fig.~\ref{aco-shape-2ab-rec}.

\begin{figure}[!ht]
\begin{center} 
{\epsfig{file=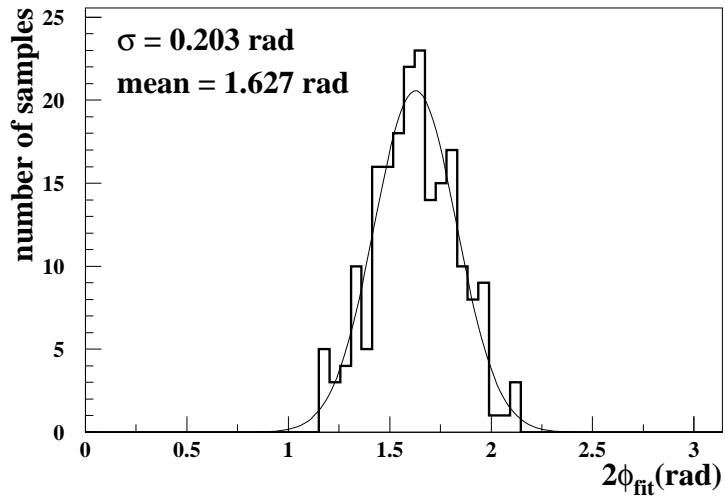,width=0.6\linewidth}}
\end{center} 
\caption  
{\it Distribution of the fit values for $2\phi$ from 400
independent samples, each corresponding to a luminosity of 1
ab$^{-1}$, for a generated value of $\phi = \pi/4$.  The curve
represents the fit of a Gaussian to this distribution. Its width
represents the expected statistical error on $2\phi$.}
\label{phase-values}
\end{figure}

In order to assess the expected accuracy and a possible experimental
bias of the $\phi$ measurement, the above procedure was repeated 400
times with acoplanarity distributions extracted from independent
samples of 1~ab$^{-1}$ luminosity each, with a nominal value of $\phi
= \pi/4$.  Unlike what was done before in
Fig.~\ref{aco-shape-2ab-rec},   the data for the two ranges of value
of $y_1 y_2$ were  appropriately combined  into one $\varphi^{*}$
distribution before the fit. The new value of $\varphi^{*}$
for the case of $y_1 y_2 < 0$ had to be redefined as $\varphi^{*}+\pi$ for 
   $0<\varphi^{*}<\pi$ and $\varphi^{*}-\pi$
for $\pi<\varphi^{*}<2\pi$. The
distribution of the fit results on $2 \phi$ for each of the
experiments  is shown in Fig.~\ref{phase-values}.  The mean value is
$1.627\pm0.014$, compared to the $\frac{\pi}{2}$  input value.  The
resulting bias of approximately 3$^{\circ}$ can probably be corrected
in the future.  The expected error on $2 \phi$ is obtained as the
width of this distribution.  It amounts to $0.20\pm 0.01$ rad, or
approximately 12$^{\circ}$.  Thus, a precision on $\phi$ of
approximately 6$^{\circ}$ can be anticipated for a SM Higgs cross
section and $h\to\tau^+\tau^-$ branching ratio at $\sqrt{s} = 350
$~GeV and 1~ab$^{-1}$. Note that so far backgrounds neither from  
other Higgs boson decays nor from other SM processes have been
considered.  While previous studies~\cite{brient:2002} have shown that
$h\to\tau^+\tau^-$ events can be selected without large backgrounds,
some small deterioration and a further lowered signal efficiency  are
to be expected. Because of  the small observed bias, it is not
expected that systematic effects will limit the resolution even for
production cross sections a few times larger than in the SM.


\section{Conclusions}

We have found that for an integrated luminosity of  1 ab$^{-1}$, at
350 GeV center-of-mass energy, a high precision LC detector such as
the proposed  TESLA, should be able to measure the
scalar--pseudoscalar mixing angle for the $h \tau \tau$ coupling with
6$^{\circ}$  accuracy in the case of a Standard Model Higgs boson  of
120 GeV mass.  The experimental error is expected to be  dominated by
statistics.

However, if the production mechanism of the Higgs boson happened to be
non-SM and  larger, the systematic errors  not studied so far and
possible new, unknown phenomena  may have some significant influence.
On the basis of  studies performed to date, we believe however  that,
if the cross section were somewhat larger than the Standard Model one
and thus the uncertainty on the mixing angle due to statistics were
not smaller than 4$^{\circ}$, we do not expect the systematic error to
be a problem.  In the case of Higgs boson scenarios predicting even
higher rates of observed $h \to \tau \tau$ samples, the issue of the
systematic error definitely needs to be re-addressed before any
conclusion on measuring the scalar--pseudoscalar mixing angle in the
$h\tau\tau$ coupling  with higher  precision can be attempted.

Finally, let us note that this method can be applied  to measure the
parity properties of other scalar particles, not necessarily only
Higgs boson(s).
                                                                               
\subsection*{Acknowledgements}
Two of us (ZW and MW) would like to thank M. Peskin for raising, two
years ago, our attention to the importance of measuring, at LC,
the parity of the Higgs boson  using the $\tau\tau$ decay channel.


\providecommand{\href}[2]{#2}

\end{document}